# Towards metal electrode interface scavenging of rare-earth scandates: A $Sc_2O_3$ and $Gd_2O_3$ study

M.A. Pampillón, P.C. Feijoo, E. San Andrés, M. Toledano-Luque, A. del Prado, A.J. Blázquez, M.L. Lucía

Departamento de Física Aplicada III (Electricidad y Electrónica), Facultad de Ciencias Físicas, Universidad Complutense de Madrid E-28040, Spain

## Abstract

Amorphous $Gd_2O_3$ and $Sc_2O_3$ thin films were deposited on Si by high-pressure sputtering (HPS). In order to reduce the uncontrolled interfacial $SiO_x$ growth, firstly a metallic film of Gd or Sc was sputtered in pure Ar plasma. Subsequently, they were in situ plasma oxidized in an $Ar/O_2$ atmosphere. For post-processing interfacial $SiO_x$ thickness reduction, three different top metal electrodes were studied: platinum, aluminum and titanium. For both dielectrics, it was found that Pt did not react with the films, while Al reacted with them forming an aluminate-like interface and, finally, Ti was effective in scavenging the $SiO_2$ interface thickness without severely compromising gate dielectric leakage.

1. ## Introduction

Some promising candidates as future high-k dielectrics for high-performance MOSFETs and also for memory applications lie within the rare-earth scandate family. In particular, GdScO is one of the most promising ternaries [1,2]. Nevertheless, before it could be integrated into an industrial process, many hurdles must be overcome, like avoiding polycrystalline growth to keep low gate leakage [3], depositing under conditions that reduces the interfacial $SiO_x$ thickness to achieve a higher effective permittivity [4], presenting a low density of interfacial or bulk traps [5], etc.

Concerning the interfacial $SiO_x$ requirement, a new approach that is gaining momentum is the use of interface "scavengers" [6,7]. This approach aims to the reduction of the low-k interfacial thickness after dielectric film deposition. In $HfO_2$ this approach has achieved extremely low equivalent oxide thickness (EOT), but its applicability to alternative dielectrics is an unexplored field.

As a preliminary step before the test of the scavenging concept on GdScO, in this work we have studied the interface scavenging on its binary oxides: $Sc_2O_3$ and $Gd_2O_3$. With this aim, MIS devices were fabricated by sputtering a metallic film (Sc or Gd) on Si, in order to prevent the exposure of the Si surface to oxygen atmosphere. Thereafter, they were in situ plasma oxidized. Finally, three different gate electrodes were deposited: Pt (noble metal, which should not react with the dielectric), Al (a metal that might react, creating an aluminate) and Ti (which is a well known oxygen scavenger [7]).

## 2. Experimental

We have obtained $Sc_2O_3$ and $Gd_2O_3$ thin films by sputtering metallic Sc and Gd targets by HPS [8] in a pure Ar plasma and, afterwards, in situ oxidizing them in a mixed $Ar/O_2$ atmosphere. The substrates were n-type Si (1 0 0) 2″ wafers with a resistivity of 1.5–5 Ω cm. Wafers were cleaned by the standard RCA procedure and, just before the introduction to the chamber, they were etched in diluted HF (1:50) for 30 s. Both targets had the maximum purity available (99.9%), with a diameter of $2^{00}$ and a thickness of 1/8″. Once the pressure was in the $10^5$ mbar range, the substrates were heated to 500 C during 5 min. Thus, they should desorb any remaining native $SiO_2$ that may have been spontaneously grown during loading in.

The first deposition step was sputtering the metallic targets in a pure Ar atmosphere during a fixed time of 60 s. Base pressure was below 8 $10^7$ mbar. Plasma conditions were 0.5 mbar of pressure, 40 sccm of Ar and a rf power of 30 W. Plasma was generated by a Huttinger 300 rf generator with an excitation frequency of 13.56 MHz.

Once the metal deposition was carried out, the films were in situ plasma oxidized in the HPS chamber for 300 s by introducing a small amount of pure oxygen, in an Ar/$O_2$ 95/5 ratio. Therefore the fluxes were 38 sccm of Ar and 2 sccm of pure $O_2$. In this step, the pressure and the rf power excitation were the same as in the previous step, 0.5 mbar and 30 W, respectively. Both processes were carried out at room temperature without heating the substrate. The optimum plasma conditions were selected by glow discharge optical spectroscopy (GDOS) measuring at wavelengths between 280 and 520 nm.

To study the crystalline structure, some samples were characterized by glancing incidence X-ray diffraction (GIXRD). The measuring system was a XPERT MRD of Panalytical in a h–2h configuration using the Cu $K_a$ line.

MIS devices were fabricated by defining square openings on top of both dielectrics with a negative resist n-LOF 2035 lithographic process. The side length of the square openings varied from 630 to 100 um. To guarantee that both dielectrics were as identical as possible for all three metals, instead of preparing a different wafer for each metal, one wafer per dielectric was cut into four quarters. Later, different top electrodes were e-beam evaporated and lift-off on each piece: Al (with a thickness of 40 nm), Pt (only 3 nm thick, to avoid adhesion problems) and Ti (30 nm). Pt and Ti were then capped with 30 nm of Al (in the case of Pt, to avoid structural degradation when probing, and for Ti, to avoid surface oxidation). After lift-off, the Ti/Al backside electrode was evaporated and a forming gas anneal (FGA) was performed for 20 min at 300 C.

The resulting C–V, G–V and J–V curves of the MIS devices were measured with an Agilent 4294A impedance meter and a Keithley 4200, respectively, close to the centre of the original wafer. In order to obtain the equivalent oxide thickness (EOT) of both dielectrics, C–V curves were fitted with Hauser's CVC algorithm [9].

Finally, the MIS samples were sliced, thinned and prepared for transmission electron microscopy (TEM). This way, the thickness information obtained from TEM can be used for discussion without ambiguity. The cross-sectional images were taken with a Tecnai T20 microscope from FEI, operating at 200 keV.

3. Results and discussion

To gain insight of the growing mechanism of the sputtered films, we studied the optical spectra of the plasma measured by GDOS. Fig. 1 shows four different GDOS spectra for pure Ar (Fig. 1 traces a and c) and mixed $Ar/O_2$ atmospheres (Fig. 1 traces b and d). In all cases, the rf power excitation was 30 W and the pressure 0.5 mbar. Comparing Fig. 1a and b it can be seen that when sputtering in an Ar plasma, Sc peaks (at 391, 402 and 474 nm) [10] are extremely intense. However, the introduction of oxygen to the chamber changes the spectra completely and those peaks disappear. This same behaviour can be observed when sputtering Gd: in pure Ar (trace c), Gd peaks (between 300 and 380 nm) [11] are quite strong, but introducing $O_2$ has the effect that no clear signal of Gd from the target is found. On traces b and d only neutral Ar peaks can be clearly identified (located between 400 and 450 nm [12]) or ionized Ar peaks (at 450–490 nm [12]), but with lower intensity than in the pure Ar case. This suggests that when introducing oxygen to the chamber no deposition takes place, or at least the deposition rate is significatively reduced.

To decide whether when oxygen is introduced only an oxidation process takes place or there is also some deposition, Fig. 2 shows the cross-sectional TEM image of a film obtained by sputtering a Gd target during 5 min with an $Ar/O_2$ plasma in the same conditions of Fig. 1. There, it can be observed that no evidence of a $Gd_2O_3$ layer is found. If it were present, it would appear as a dark layer between the glue and Si. In this figure, only an amorphous bright layer 3 nm thick over Si can be found. Together with GDOS results, the most likely explanation is that at this low rf power there is no Gd extraction.

Thus, the bare Si substrate is exposed to the excited Ar/O$_2$ atmosphere, resulting on the oxidation of the substrate and creating a SiO$_x$ film. In other words, when the power is 30 W instead of sputtering, we are only oxidizing the sample. This oxidation is effective even at room temperature, as the TEM image proves.

In order to further study the plasma oxidation conditions, we sputtered on some bare Si substrates for long times (30–90 min) at several rf powers in the Ar/O$_2$ atmosphere. The GIXRD results are presented in Fig. 3. On traces a–c the metallic target is Gd, and in trace d is metallic Sc. For Gd we present several rf powers: 30, 40 and 50 W. We can see that the substrates sputtered with Gd for plasma powers above of 40 W present the monoclinic Gd$_2$O$_3$ phase [13]; nevertheless, at 30 W this does not happen, because only the substrate peak corresponding to the Si(3 1 1) plane is observed. On the other hand, in the spectrum of Sc at 50 W (trace d), no diffraction can be seen, and neither at lower rf powers (not shown). These results can be explained as follows: for gadolinium targets and rf powers higher or equal than 40 W, there is not only oxidation but also deposition. However, at rf power of 30 W, no gadolinium oxide is deposited, and only an oxidation process is taking place. For scandium this effect is more pronounced, since there is no scandium oxide deposition even at 50 W.

Fig. 4 shows representative C–V curves of MIS devices for the three different top metal electrodes studied. For both dielectrics (Fig. 4a shows the results with Sc$_2$O$_3$ as a dielectric and Fig. 4b for Gd$_2$O$_3$) samples evaporated with Al have the lowest capacitance (therefore, highest EOT). This is possibly due to the formation of aluminates. When Pt is used as top electrode, the EOT is about 0.4–0.5 nm smaller than Al. The explication of this is that Pt is a noble metal and does not react with the dielectric. For this reason, Pt is useful to study the plain properties of dielectric films. Finally, Ti presents, by far, the highest capacitance: lowest EOT (near 2 nm less than Pt), and 1.75 nm of EOT for Sc$_2$O$_3$ and 1.8 nm for Gd$_2$O$_3$. The explanation is that the Ti should be working as a scavenger of the interfacial SiO$_x$ [7].

The conductance peak in depletion can be used to estimate interface trap density ($D_{it}$). As opposed to the total capacitance results, the $D_{it}$ trend is not so positive for Ti devices: the $D_{it}$ values are degraded for Ti (2-4 $10^{12}$ eV$^1$ cm$^2$) compared to Al and Pt devices with show a good quality value of 1-1.5 $10^{11}$ eV$^1$ cm$^2$. In Pt and Al during the FGA hydrogen diffuses from the forming gas to the interface and passivates dangling bonds. On the other hand, since Ti is scavenging oxygen from the interface, so it may create extra dangling bonds that produce an increase in trap density. This implies that further optimization would be needed to improve the interface. Also, concerning the flatband voltage, there is no clear trend (ideally Ti an Al should have a similar $V_{fb}$, 1 V lower than Pt [7]). The most likely origin of this result is the different $D_{it}$ and/or the bulk density of defects, which also produce a shift on the $V_{fb}$.

To investigate the large differences observed in the capacitance value for both dielectrics, TEM images were obtained for the $Gd_2O_3$ films. On the top of Fig. 5, we show the image for Pt capped with Al as metal electrode. It can be seen that the interfacial $SiO_x$ thickness is about 1.8 nm under 3.1 nm of an amorphous $Gd_2O_3$ layer. The second image of Fig. 5 shows that for Al devices the thicknesses of the $SiO_x$ and $Gd_2O_3$ films are identical, but they also present an amorphous 2.7 nm film on top of the $Gd_2O_3$ layer. This could be explained as a formation of an aluminate-like film at the metal/ high-k interface, that can justify the increase in EOT found. Finally, the Ti capacitors have a much thinner $SiO_x$ interface, less than 1 nm thick. The reason of this reduction must be the scavenging of the $SiO_x$ interface by the Ti metal electrode. It is noteworthy that all these pictures show an amorphous $Gd_2O_3$ film, which for high-k applications is a very good result.

Finally, we show in Fig. 6 the J–V results in accumulation for MIS devices with three different top metal electrodes. Fig. 6a presents curves for $Sc_2O_3$ as a dielectric, while Fig. 6b for $Gd_2O_3$. Despite the significant differences in the EOT observed previously for the three metals, we see that the leakage current is not unduly compromised. The three curves are comparable for both dielectrics. In fact, quite surprisingly, for the Ti/Gd$_2$O$_3$ case the leakage is lower than with the other two metals. A similar effect was also found in Ref. [7].

4. Conclusions

In this work we found that the interface scavenging concept can be also applied for the optimization of Sc$_2$O$_3$ and Gd$_2$O$_3$. This is a promising result towards the application of interfacial SiO$_2$ scavenging to rare-earth scandates. Our future work will focus on this study. Acknowledgements

The authors would like to acknowledge 'C.A.I. de Técnicas Físicas' and 'C.A.I. de Rayos-X' of the Universidad Complutense de Madrid, and 'Instituto de Nanoelectrónica de Aragón'. This work was funded by the project TEC2010-18051. P.C. Feijoo work was funded by MICINN through the FPU Grant AP2007-01157.

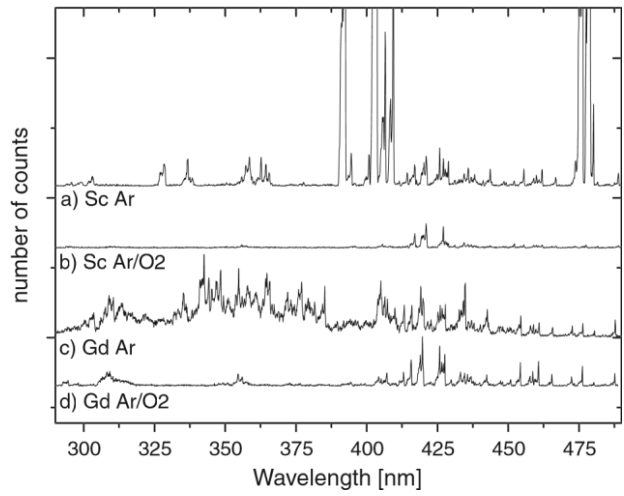

Fig. 1. Optical spectra of Sc and Gd targets sputtered in pure Ar and Ar/$O_2$ 95/5 mixed atmospheres. For all spectra, rf power is 30 W and pressure is 0.5 mbar.

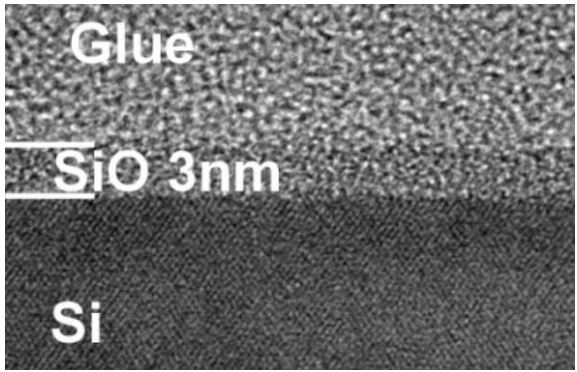

Fig. 2. Cross-sectional TEM images of a bare Si substrate exposed to an Ar/O$_2$ 30 W plasma during 5 min at room temperature.

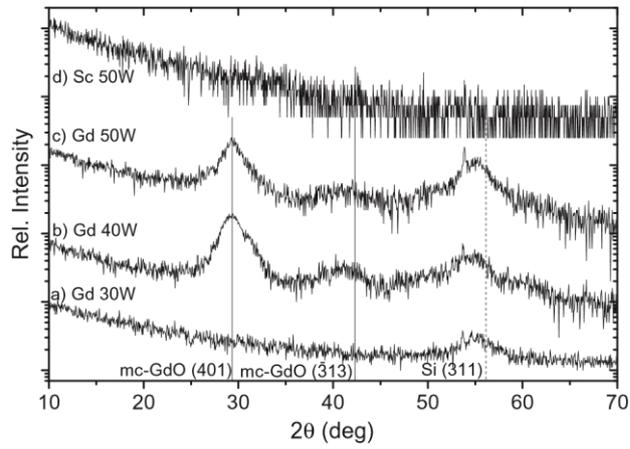

Fig. 3. Glancing incidence X-ray diffraction spectra of the substrates sputtered in Ar/O$_2$ atmosphere: (a) Gd target 30 W of rf power, (b) Gd 40 W, (c) Gd 50 W and (d) Sc target 50 W.

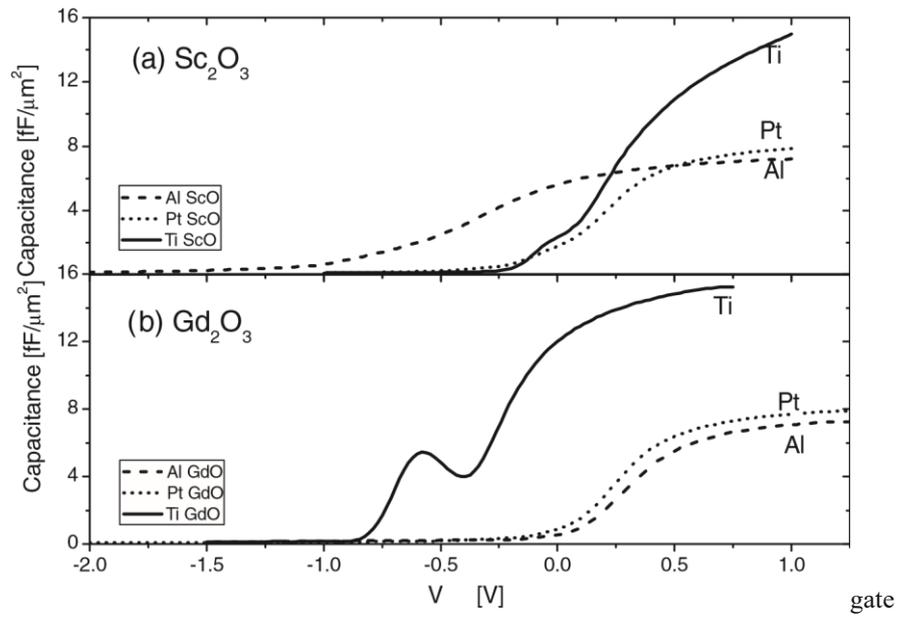

Fig. 4. Representative gate capacitance curves as function of gate voltage measured at 10 kHz for different metal electrodes (Al, Pt and Ti): (a) $Sc_2O_3$ and (b) $Gd_2O_3$. All the

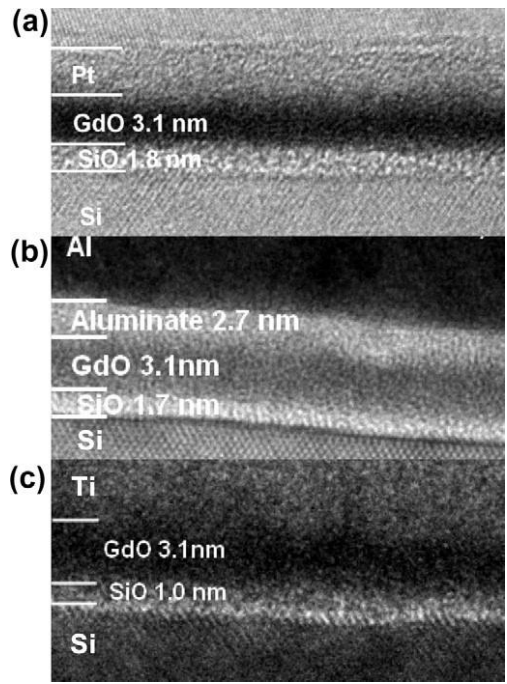

Fig. 5. Cross-sectional TEM images of the $Gd_2O_3$ MIS devices forming gas annealed at 300 C for 20 min with different gate electrodes: (a) Pt, (b) Al and (c) Ti.

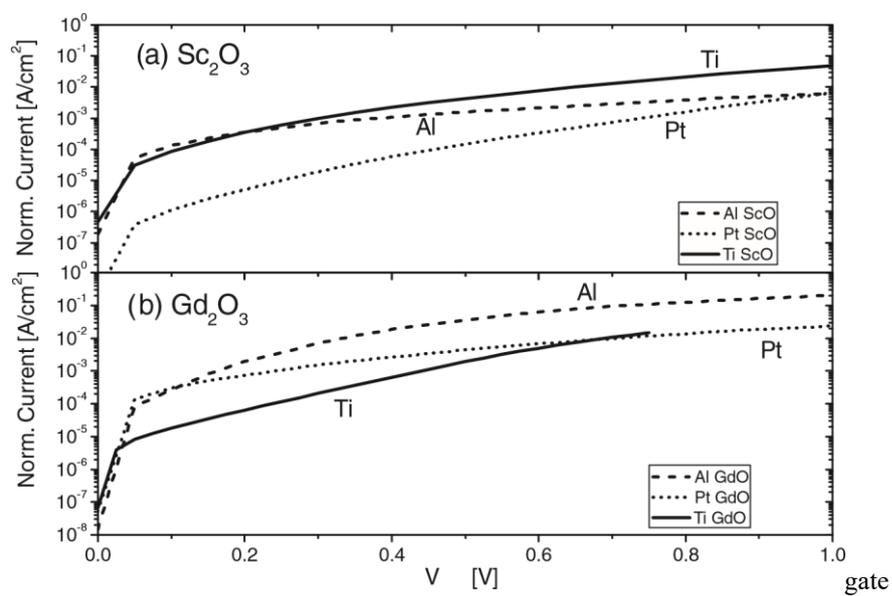

Fig. 6. Representative gate current density as a function of gate voltage in accumulation for different metal electrodes (Al, Pt and Ti) after annealing with forming gas for 20 min ant 300 C: (a) $Sc_2O_3$ and (b) $Gd_2O_3$.